\documentclass[12pt]{article}
\usepackage{amssymb,latexsym,amsmath}

\usepackage{color}
\usepackage{rotating}
\usepackage{arydshln}

\usepackage{epsfig,lscape}
\usepackage{color}

\makeatletter
\@ifundefined{MPFourScale}{\def\MPFourScale{1.00}}{}
\DeclareFontFamily{U}{mp4}{}%
\DeclareFontShape{U}{mp4}{m}{n}{<->s * [\MPFourScale]cmb10}{}
\DeclareSymbolFont{boldgreekuc}{U}{mp4}{m}{n}
\DeclareMathSymbol{\bfAlpha}{\mathord}{boldgreekuc}{"41}
\DeclareMathSymbol{\bfBeta}{\mathord}{boldgreekuc}{"42}
\DeclareMathSymbol{\bfPsi}{\mathord}{boldgreekuc}{"09}
\DeclareMathSymbol{\bfDelta}{\mathord}{boldgreekuc}{"01}
\DeclareMathSymbol{\bfEpsilon}{\mathord}{boldgreekuc}{"45}
\DeclareMathSymbol{\bfPhi}{\mathord}{boldgreekuc}{"08}
\DeclareMathSymbol{\bfGamma}{\mathord}{boldgreekuc}{"00}
\DeclareMathSymbol{\bfEta}{\mathord}{boldgreekuc}{"48}
\DeclareMathSymbol{\bfIota}{\mathord}{boldgreekuc}{"49}
\DeclareMathSymbol{\bfXi}{\mathord}{boldgreekuc}{"04}
\DeclareMathSymbol{\bfKappa}{\mathord}{boldgreekuc}{"4B}
\DeclareMathSymbol{\bfLambda}{\mathord}{boldgreekuc}{"03}
\DeclareMathSymbol{\bfMu}{\mathord}{boldgreekuc}{"4D}
\DeclareMathSymbol{\bfNu}{\mathord}{boldgreekuc}{"4E}
\DeclareMathSymbol{\bfPi}{\mathord}{boldgreekuc}{"05}
\DeclareMathSymbol{\bfTheta}{\mathord}{boldgreekuc}{"02}
\DeclareMathSymbol{\bfRho}{\mathord}{boldgreekuc}{"52}
\DeclareMathSymbol{\bfSigma}{\mathord}{boldgreekuc}{"06}
\DeclareMathSymbol{\bfTau}{\mathord}{boldgreekuc}{"54}
\DeclareMathSymbol{\bfVartheta}{\mathord}{boldgreekuc}{"02} 
\DeclareMathSymbol{\bfOmega}{\mathord}{boldgreekuc}{"0A}
\DeclareMathSymbol{\bfVarphi}{\mathord}{boldgreekuc}{"08} 
\DeclareMathSymbol{\bfUpsilon}{\mathord}{boldgreekuc}{"07}
\DeclareMathSymbol{\bfZeta}{\mathord}{boldgreekuc}{"5A}

\DeclareFontFamily{U}{mp4sl}{}%
\DeclareFontShape{U}{mp4sl}{m}{n}{<->s * [\MPFourScale]cmmib10}{}
\DeclareSymbolFont{boldgreek}{U}{mp4sl}{m}{n}
\DeclareMathSymbol{\bfalpha}{\mathord}{boldgreek}{"0B}
\DeclareMathSymbol{\bfbeta}{\mathord}{boldgreek}{"0C}
\DeclareMathSymbol{\bfpsi}{\mathord}{boldgreek}{"20}
\DeclareMathSymbol{\bfdelta}{\mathord}{boldgreek}{"0E}
\DeclareMathSymbol{\bfepsilon}{\mathord}{boldgreek}{"0F}
\DeclareMathSymbol{\bfphi}{\mathord}{boldgreek}{"1E}
\DeclareMathSymbol{\bfgamma}{\mathord}{boldgreek}{"0D}
\DeclareMathSymbol{\bfeta}{\mathord}{boldgreek}{"11}
\DeclareMathSymbol{\bfiota}{\mathord}{boldgreek}{"13}
\DeclareMathSymbol{\bfxi}{\mathord}{boldgreek}{"18}
\DeclareMathSymbol{\bfkappa}{\mathord}{boldgreek}{"14}
\DeclareMathSymbol{\bflambda}{\mathord}{boldgreek}{"15}
\DeclareMathSymbol{\bfmu}{\mathord}{boldgreek}{"16}
\DeclareMathSymbol{\bfnu}{\mathord}{boldgreek}{"17}
\DeclareMathSymbol{\bfpi}{\mathord}{boldgreek}{"19}
\DeclareMathSymbol{\bfvartheta}{\mathord}{boldgreek}{"23}
\DeclareMathSymbol{\bfrho}{\mathord}{boldgreek}{"1A}
\DeclareMathSymbol{\bfsigma}{\mathord}{boldgreek}{"1B}
\DeclareMathSymbol{\bftau}{\mathord}{boldgreek}{"1C}
\DeclareMathSymbol{\bftheta}{\mathord}{boldgreek}{"12}
\DeclareMathSymbol{\bfomega}{\mathord}{boldgreek}{"21}
\DeclareMathSymbol{\bfvarphi}{\mathord}{boldgreek}{"27}
\DeclareMathSymbol{\bfchi}{\mathord}{boldgreek}{"1F}
\DeclareMathSymbol{\bfupsilon}{\mathord}{boldgreek}{"1D}
\DeclareMathSymbol{\bfzeta}{\mathord}{boldgreek}{"10}

\makeatother

\begin{document}
\title{A pairwise likelihood approach to simultaneous clustering and dimensional reduction of ordinal data}

\author{Monia Ranalli\thanks{Department of Statistics, The Pennsylvania State University, USA {\tt monia.ranalli@psu.edu}} \and  \and Roberto Rocci\thanks{IGF Department, University of Tor Vergata, Rome, {\tt roberto.rocci@uniroma2.it}}}

\date{}

\maketitle

\begin{abstract}
The literature on clustering for continuous data is rich and wide; differently, that one developed for categorical data is still limited. In some cases, the problem is made more difficult by the presence of noise variables/dimensions that do not contain information about the clustering structure and could mask it. The aim of this paper is to propose a model for simultaneous
clustering and dimensionality reduction of ordered categorical data able to detect the discriminative dimensions discarding the noise ones. Following the underlying response variable approach, the observed variables are considered as a discretization of underlying first-order latent continuous variables distributed as a Gaussian mixture. To recognize discriminative and noise dimensions, these variables are considered to be linear combinations of two independent sets of second-order latent variables where only one contains the information about the cluster
structure while the other contains noise dimensions. The model specification involves
multidimensional integrals that make the maximum likelihood estimation cumbersome and in some cases infeasible. To overcome this issue the parameter estimation is carried out through an EM-like algorithm maximizing a pairwise log-likelihood. Examples of application of the model on real and simulated data are performed to show the effectiveness of the proposal.
\\

\textbf{Keywords}:Mixture models, Reduction data, Ordinal data, Pairwise Likelihood, EM algorithm

\end{abstract}\vspace{\fill}\pagebreak

\vspace{\fill}\newpage

\section{Introduction}
Cluster analysis aims at partitioning the data into meaningful groups which should differ considerably from each other. The literature on clustering for continuous data is rich and wide; differently, that one developed for categorical data is still limited. In fact, only in the last decades there has been an increasing interest in clustering categorical data, although they are encountered in many fields, such
as in behavioural, social and health sciences. These variables are frequently of ordinal type, measuring attitudes, abilities or opinions, and practitioners often apply on their ranks models and techniques developed for continuous data. Several authors have shown how this procedure can give biased estimates and is definitely less efficient than a proper modelization that is able to take into account the ordinal nature of the data (e.g. \cite{mr}). Such models mainly adopt two approaches developed in factor analysis framework: IRT (Item Response Theory) and URV (Underlying Response Variables). In the former, the probabilities of the categories are assumed to be analytic functions of some latent variables having a particular cluster structure. The best known model is latent class analysis (LCA; \cite{goodman74}) where the latent variable is nominal. Examples where the latent variables are continuous are found in \cite{viroli12}, \cite{mcparland2012}, \cite{gollini2012}. In the URV approach, the ordinal variables are seen as a discretization of continuous latent variables jointly distributed as a finite mixture; examples are: \cite{everitt88}, \cite{lubke08}, \cite{mr}. In both approaches, the use of latent continuous variables makes the estimation rather complex because it requires the computation of many high dimensional integrals. The problem is usually solved by approximating the log-likelihood function. Indeed several lines of research propose different approximations, but they share the same idea: replacing the full likelihood with a surrogate that is easier to maximize and make
inference about model parameters. In this regard we mention some useful surrogate functions, such as the variational likelihood \cite{gollini2012,tang2014} or the pairwise likelihood  \cite{mr} to cluster categorical or ordinal data, respectively. Beside this, other approaches based on simulating the hidden variables exist.
\\
In some cases, the clustering problem is made more difficult by the presence of variables and/or dimensions (named noise) that are uninformative for recovering the latent groups and could obscure the true cluster structure. Different approaches exist in literature to identify discriminative dimensions that emphasize group separability and give a representation of the cluster structure discarding the irrelevant and redundant noise dimensions. We can distinguish between variable selection and dimensionality reduction approaches.\\
In the first we find proposals which aims at estimating the cluster pattern by selecting the set of variables which best describes the cluster structure. In the context of continuous data, \cite{raftery06} formulates the problem of variable selection as a model comparison problem using the BIC, in which the variables are partitioned into two exclusive subsets representing the relevant, or discriminative, and the irrelevant, or noise, variables, respectively. \cite{maugis09} extend this approach, while \cite{witten2010} propose to perform the variable selection by using a lasso-penalty. Many other authors have extended the aforementioned works or proposed different approaches but almost exclusively on continuous data. In the context of categorical data there are only few proposals. We mention \cite{dean2010} and \cite{white2014} who extend the work of \cite{raftery06} to the latent class model.\\
On the other hand, the dimensionality reduction approach aims at discarding the irrelevant dimensions by identifying a reduced number of latent variables containing the information about the cluster structure. The easiest way to implement this approach is the so-called tandem analysis \cite{arabie94}. It is a two step procedure, where in the second step a clustering model/method is applied on a reduced number of dimensions identified in the first step. Depending on the scale measurement of the data, the first step can be implemented by using either principal components analysis (PCA), factor analysis, PCA for qualitative data \cite{young78} or multiple correspondance analysis (\cite{greenacre84}). Of course, it is difficult to find the discriminative dimensions without knowing the cluster structure. In fact, the main problem involved by tandem is that there is no guarantee that the reduced data obtained in step one is optimal for recovering the cluster structure in step two (\cite{chang83} \cite{arabie94}). This may hide or even distort the cluster structure underlying the data. As a solution to the problem, data reduction and clustering analysis should be performed simultaneously. In this way the latent factors are identified to highlight the cluster structure rather than, as happens in some cases, to obscure it. Several techniques for simultaneous clustering and dimensionality reduction (SCR) have been proposed in a non-model based framework for quantitative (e.g.: \cite{vichi2001}; \cite{rocci2011}) or categorical data (e.g.: \cite{van89}; \cite{hwang06}). \\
There are also approaches based on a family of mixture models which fit the data into a common discriminative subspace (see e.g. \cite{kumar98,bouveyron12b}). The key idea is to assume a common latent subspace to all groups that is the most discriminative one. This allows to project the data into a lower dimensional space preserving the clustering characteristics in order to improve visualization and interpretation of the underlying structure of the data. The model can be formulated as a finite mixture of Gaussians with a particular set of constraints on the parameters.\\
It is worth pointing out that SCR partially overlaps with the parsimony criterion. 
Indeed in high/dimensional context, the curse of dimensionality lead to define models capturing the essential clustering features reducing the number of parameters. One of the earliest \textit{parsimonious} proposal is given by the mixture of factor analyzers (MFA). The
MFA model differs from the factor analysis model in having different local factor models. Conversely, the standard factor analysis assumes a common factor model. The MFA to cluster the data and reduce locally the dimensionality of each cluster simultaneously  was originally proposed
by \cite{ghahramani97} and \cite{hinton97}. Later, a general framework for the MFA model was proposed by \cite{celeux95,mcnicholas2008}. Furthermore, we point the reader to see also \cite{tipping99} and \cite{bishop98} who considered
the related model of mixtures of principal component analyzers for the same purpose.  Further references may be found in chapter 8 of \cite{peel00} and in a recent review on model-based clustering of high-dimensional data \cite{bouveyron12}. As regards categorical data, we find few analogous proposals (see e.g. \cite{gollini2012,mcparland2012,marbac14,marbac14b}).\\

The aim of this paper is to propose a model for SCR on ordered categorical data. Following the URV approach, the observed variables are considered as a discretization of underlying first-order latent continuous variables. To detect noise dimensions, the latent variables are considered to be linear combinations of two independent sets of second-order latent variables where only one contains the information about the cluster structure, defining a discriminative subspace, while the other one contains noise dimensions. Technically, the variables in the first set are distributed as a finite mixture of Gaussians while in the second set as a multivariate normal. It is important to note that when in the dataset there are noise variables then they tend to coincide with the set of second order noise latent variables. If they are not present then the model could be still able to identify a reduced set of second order discriminative latent dimensions. This allow us to reduce the number of parameters and identify the main features of the clustering structure. The model specification involves multidimensional integrals that make the maximum likelihood estimation rather cumbersome and in some cases infeasible. To overcome this issue, the model is estimated within the EM framework maximizing the pairwise log-likelihood, i.e. the sum of all possible log-likelihoods based on the bivariate marginals, as proposed in \cite{mr}. The estimators obtained have been proven to be consistent, asymptotically unbiased and normally distributed. In general they are less efficient than the full maximum likelihood estimators, even if in many cases the loss in efficiency is very small or almost null \cite{lindsay88,varin2011}, but much more efficient in terms of computational complexity. 
\\
The plan of the paper is the following: in the second section we present the model; in section 3 we describe how to take into account the presence of noise dimensions and/or variables; then the pairwise algorithm used to estimate the model parameters is presented in section 4. Section 5, 6 and 7 deal with model identifiability issue, the output interpretation and the model selection problem, respectively. In section 8 a simulation study has been conducted to investigate the behaviour of the proposed methodology, while in section 9 an application to real data is shown. In the last section some remarks are pointed out.
\section{Model}
Let $x_{1},x_{2},\ldots ,x_{P}$ be ordinal variables and $c_{i}=1,\ldots , C_{i}$ the associated categories for $i=1,2,\ldots , P$. There are $R=\prod_{i=1}^{P}C_{i}$ possible response patterns, which have the following form $ \textbf{x}_{r}=(x_{1}=c_{1},x_{2}=c_{2},\ldots ,x_{P}=c_{P})$.
Let $\bf{y}$ be the heteroscedastic latent Gaussian mixture $f\left( \bf{y}\right) =\sum_{g=1}^{G}p_{g}\phi\left(\bf{y}; \bfmu_{g},\bfSigma_{g}\right)$, where the $p_g$'s are the mixing weights and $\phi\left(\bf{y}; \bfmu_{g},\bfSigma_{g}\right)$ is the density of a $P$-variate normal distribution with mean vector $\bfmu_g$ and covariance matrix $\bfSigma_g$. Under the URV approach, the ordinal variables are considered as a discretization of $\textbf{y}$, i.e. generated by thresholding $\textbf{y}$, as follows,
\begin{center}
$\gamma _{c_i-1}^{(i)} \leq y_{i} < \gamma _{c_{i}}^{(i)} \Leftrightarrow  x_{i}=c_{i}$,
\end{center}
where $-{\large \infty}  =\gamma _{0}^{(i)}< \gamma _{{1}}^{(i)}<\ldots < \gamma _{{C_i-1}}^{(i)}< \gamma _{{C_i}}^{(i)}=+{\large \infty} $ are the thresholds defining the $C_i$ categories.

Let us set $\bfpsi=\left\lbrace \textit{p}_1,\ldots,\textit{p}_{G},\bfmu_1,\ldots,\bfmu_G,\bfSigma_1,\ldots,\bfSigma_G,\bfgamma \right\rbrace \in \bfPsi$, where $\bfPsi$ is the parameter space.
The probability of response pattern $\textbf{x}_r$ is given by
\begin{eqnarray*}
Pr(x_{1}=c_{1},x_{2}=c_{2},\ldots ,x_{P}=c_{P};\bfpsi)&=&\sum_{g=1}^{G}\textit{p}_{g}\int_{\gamma _{c_1-1}^{(1)}}^{\gamma _{c_{1}}^{(1)}}\cdots \int_{\gamma _{c_P-1}^{(P)}}^{\gamma _{c_{P}}^{(P)}} \phi(\textbf{y};\bfmu_{g},\bfSigma_{g})d\textbf{y} \\
&=&\sum_{g=1}^{G}\textit{p}_g\pi_{r}(\bfmu_{g},\bfSigma_{g},\bfgamma), 
\end{eqnarray*}

where $\pi_r\left(\bfmu_g,\bfSigma_g,\bfgamma\right) $ is the probability of response pattern $\textbf{x}_r$ in the cluster $g$ and $p_g$ is the probability of belonging to group $g$ subject to $p_g>0$ and $\sum_{g=1}^{G}p_g=1$.

 Thus, for a random i.i.d. sample of size $N$ the log-likelihood is
\begin{eqnarray}
\ell(\bfpsi;\textbf{x})&=&\sum_{r=1}^{R}\textit{n}_{r}\log\left[ \sum_{g=1}^{G}\textit{p}_{g}\pi_{r}\left(\bfmu_{g},\bfSigma_{g},\bfgamma\right)\right], 
\end{eqnarray}
where $n_r$ is the observed sample frequency of response pattern $r$ and $\sum_{r=1}^{R} n_r=N$.

\section{How to detect the presence of noise variables}
Sometimes noisy dimensions are present in the data.
These are dimensions that do not contain information about the
cluster structure and could mask the true classes.
It means that there exists a proper discriminative subspace, 
with a dimension less than the number of variables, where the clusters lie.
In order to identify the discriminative subspace, in the previously described
model it is assumed that there is a second order set of $P$ latent variables $\tilde{\textbf{y}}$, which in turn is formed of two independent subsets of variables. In the first there are $Q$ (with $Q<P$) variables that have some clustering information, while in the second set there are $P-Q$ noisy variables. Thus, it is assumed that only the first $Q$ elements of $\tilde{\textbf{y}}$ carry any class discrimination information defining the so-called discriminative subspace. Technically, the $Q$ informative elements are assumed to be distributed as a
mixture of Gaussians with class conditional means and variances equal to 
$E(\tilde{\textbf{y}}^Q\mid g)=\bfeta_g$ and $\text{Cov}(\tilde{\textbf{y}}^Q\mid g)=\bfOmega_g$. The $P-Q$ noisy elements do not have 
information about the cluster structure, it follows that they are independent of 
$\tilde{\textbf{y}}^Q$ and their distribution 
does not vary from one class to another. In particular we assume that
$E(\tilde{\textbf{y}}^{\bar{Q}}\mid g)=\bfeta_0$ and $\text{Cov}(\tilde{\textbf{y}}^{\bar{Q}}\mid g)=\bfOmega_0$. The link between the two orders of latent variables $\tilde{\textbf{y}}$ and $\textbf{y}$ is given by a non-singular $P \times P$ matrix  $\textbf{A}$, as $\textbf{y}=\textbf{A}\tilde{\textbf{y}}$.
This means requiring a particular structure on the mean vectors and covariance matrices of $\textbf{y}$. The assumption of multivariate normality in each component provides a convenient way of specifying the parameter structure.
For each component $g$, the mean vector and the covariance matrix have the following structures,
\begin{equation*}
\bfmu_{g}=E(\textbf{y}\mid g)=\textbf{A}E(\tilde{\textbf{y}}\mid g)=\textbf{A}\begin{bmatrix}
\eta_{g,1} \\ 
\vdots \\ 
\eta_{g,Q}  \\ 
\eta_{0,1}  \\ 
\vdots\\ 
\eta_{0,P-Q}
\end{bmatrix}=\textbf{A}\begin{bmatrix}\bfeta_{g}\\
\bfeta_0
\end{bmatrix}
\end{equation*}

and
\begin{equation*}
\bfSigma_{g}=\text{Cov}(\textbf{y}\mid g)=\textbf{A}\text{Cov}(\tilde{\textbf{y}}\mid g)\textbf{A}^{\prime}=\textbf{A}\begin{bmatrix}
\bfOmega_{g}& \textbf{0}\\
\textbf{0} & \bfOmega_0
\end{bmatrix} \textbf{A}^{\prime}.
\end{equation*}

%
%
%
%
\section{Pairwise EM algorithm}
In the previous section we have seen how to reparametrize  the model described in section 2 in order to identify discriminative/noise dimensions. An efficient way to estimate it would be through the maximization of the likelihood. However, the likelihood function involves
multidimensional integrals, whose evaluation is computationally demanding as the
number of observed variables increases. 
Indeed  multidimensional integrals should be evaluated for each response pattern in the sample at several points of the parameter space. Thus the model estimation through a full
maximum likelihood approach becomes prohibitive with $P$ greater than 5 and still demanding with a very low number of variables $P$. As suggested in \cite{mr}, the model is estimated within the expectation-maximization (EM) framework maximizing a pairwise likelihood. 
It is a robust estimation method and its estimators have been proven to be consistent,
asymptotically unbiased and normally distributed, under regularity conditions \cite{lindsay88,varin2011,molenberghs2005}. In general they are less efficient
than the full maximum likelihood estimators, but in many cases the loss in efficiency
is very small or almost null \cite{lindsay88,mardia09}.

The pairwise log-likelihood is
\begin{small}
\begin{equation}
p\ell(\bfpsi;\textbf{x})=\sum_{i=1}^{P-1}\sum_{j=i+1}^{P}\ell(\bfpsi;(x_{i},x_{j}))=\sum_{i=1}^{P-1}\sum_{j=i+1}^{P}\sum_{c_{i}=1}^{C_{i}}\sum_{c_{j}=1}^{C_{j}}n_{c_{i}c_{j}}^{(ij)}\log\left[ \sum_{g=1}^{G}p_{g}\pi_{c_{i}c_{j}}^{(ij)}(\bfmu_{g},\bfSigma_g,\bfgamma)\right],
\end{equation}
\end{small}
where now, after the reparameterization, the set of models parameters is $\bfpsi=\left\lbrace \textit{p}_1,\ldots,\textit{p}_{G},\bfeta_0,\bfeta_1,\ldots,\bfeta_G,\bfOmega_0,\bfOmega_1,\ldots,\bfOmega_G,\textbf{A},\bfgamma \right\rbrace$, $n_{c_{i}c_{j}}^{(ij)}$ is the observed frequency of a response in category $c_{i}$ and $c_{j}$ for variables $x_{i}$ and $x_{j}$ respectively, while $\pi_{c_{i}c_{j}}^{(ij)}(\bfmu_{g},\bfSigma_g, \bfgamma)$ is the corresponding probability obtained by integrating the $(i,j)$ bivariate marginal of the normal distribution with parameters $\left(\bfmu_g, \bfSigma_{g}\right)$ between the given thresholds.

Let \textbf{Z} denote the group membership matrix of order {\footnotesize$ \left( \sum_{i=1}^{P-1}\sum_{j=i+1}^{P}C_{i}\times C_{j}\right) \times G$}, where $z_{c_{i}c_{j};g}^{(ij)}=1$ if the cell $(c_{i},c_{j})$ belongs to component $g$ and $z_{c_{i}c_{j};g}^{(ij)}=0$ otherwise, for $g=1,\ldots ,G$. The complete pairwise log-likelihood is

\begin{equation*}
p\ell_{c}(\bfpsi;\textbf{z},\textbf{x})=\sum_{i=1}^{P-1}\sum_{j=i+1}^{P}\sum_{c_{i}=1}^{C_{i}}
\sum_{c_{j}=1}^{C_{j}}\sum_{g=1}^{G}n_{c_{i}c_{j}}^{(ij)} z_{c_{i}c_{j};g}^{(ij)}\left[ \log\left( \pi_{c_{i}c_{j}}^{(ij)}(\bfmu_{g},\bfSigma_{g},\bfgamma)\right)+\log\left( p_{g}\right)\right] . 
\end{equation*}
The E-step requires the computation of the expected value of the complete-data pairwise log-likelihood given the current estimates of the model parameters. This is given by 
\begin{footnotesize}
\begin{eqnarray}
Q\left(\bfpsi \vert \hat {\bfpsi}^{(t-1)}  \right)&=&E_{\hat{\bfpsi}^{(t-1)}}\left[p\ell_{c}(\bfpsi;\textbf{z},\textbf{x} \vert \textbf{x}) \right]	\nonumber \\
&=&\sum_{i<j}\sum_{c_{i}=1}^{C_{i}}
\sum_{c_{j}=1}^{C_{j}}\sum_{g=1}^{G}n_{c_{i}c_{j}}^{(ij)}\hat{z}_{c_{i}c_{j};g}^{(ij)(t)}\left[ \log\left( \pi_{c_{i}c_{j}}^{(ij)}(\bfmu_{g},\bfSigma_g,\bfgamma)\right)+\log\left( p_{g}\right)\right],
\end{eqnarray}
\end{footnotesize}
where
\begin{align*}
\hat{z}_{c_{i}c_{j};g}^{(ij)(t)}=E_{\hat{{\bfpsi}}^{(t-1)}}\left[
Z_{c_{i}c_{j};g}^{(ij)}=1\vert x_i=c_i, x_j=c_j \right]=Pr_{\hat{\bfpsi}^{(t-1)}}\left[Z_{c_{i}c_{j};g}^{(ij)}=1 \vert x_i=c_i, x_j=c_j\right].  
\end{align*}

In the M-step we maximize the complete pairwise log-likelihood function subject to some constraints that will be specified in the sequel. The previous expected value is maximized with respect to the model parameters. Looking at the expected value in (3), the maximization can be decomposed in two parts: the former corresponds to the component parameters $\left(\bfmu_{g}, \bfSigma_g \right) $ and thresholds $\bfgamma$, the second one to the mixture weights $p_g$'s. The first part of the M-step has not a closed form; hence, to obtain the estimates, its maximization has been implemented in Matlab by using the command ``fmincon''  \cite{matlab} under some constraints that are explained in detail in section 5.

On the other hand, the estimate of component weight $\hat{p}_g$ has a closed form and they are easily carried out as follows,
\begin{align*}
\hat{p}_{g}=\dfrac{\sum_{i<j}\sum_{c_{i}=1}^{C_{i}}\sum_{c_{j}=1}^{C_{j}}n_{c_{i}c_{j}}^{(ij)}\hat{z}_{c_{i}c_{j};g}^{(ij)(t)}}{N},
\end{align*}
with $g=1,\cdots, G$.

In order to ensure the positive-definiteness of the covariance matrices we estimate them through their Cholesky decomposition.
It means that the objective function is maximized with respect to $\textbf{T}_g$ rather than $\bfOmega_g$ where the $\textbf{T}$s are upper triangular matrices such that $\textbf{T}_g\textbf{T}_g^{\prime}=\bfOmega_g$ for $g=0,1,\ldots,G$. Finally, the threshold parameters do not change over the components, but each component is characterised by a different set of parameters; now, standardizing each component by making a change of variable, i.e. $z_i=(y_i-\mu_{g}^{(i)})/ \sigma_g^{(ii)}$, we obtain new integration limits changing over the components. These are defined as
\begin{equation*}
\tau_{c_{i;g}}^{(i)}=\dfrac{\gamma_{c_{i}}^{(i)}-\mu_{g}^{(i)}}{\sigma_{g}^{(ii)}}.
\end{equation*}
This allows to compute the probability of a response in category $c_i$ and $c_j$ for variables $x_i$ and $x_j$, respectively, in (3) as
\begin{footnotesize}
\begin{eqnarray}
&&\pi_{c_{i}c_{j}}^{(ij)}(\textbf{0},\textbf{R}_g,\bftau_{\cdot,g}^{(i)},\bftau_{\cdot,g}^{(j)})=\int_{\tau_{c_{i}-1,g}^{(i)}}^{\tau _{c_{i},g}^{(i)}} \int_{\tau_{c_{j}-1,g}^{(j)}}^{\tau _{c_{j},g}^{(j)}}\phi\left( z_{i},z_{j};0,0,1,1,\rho_{ij}^{(g)}\right) dz_{i}dz_{j}
\\
&&=\Phi_{2}(\tau_{c_{i},g}^{(i)},\tau_{c_{j},g}^{(j)};\rho_{ij}^{(g)})-\Phi_{2}(\tau_{c_{i},g}^{(i)},\tau_{c_{j}-1,g}^{(j)};\rho_{ij}^{(g)})+ \nonumber \\
&&-\Phi_{2}(\tau_{c_{i}-1,g}^{(i)},\tau_{c_{j},g}^{(j)};\rho_{ij}^{(g)})+\Phi_{2}(\tau_{c_{i}-1,g}^{(i)},\tau_{c_{j}-1,g}^{(j)};\rho_{ij}^{(g)}),
\nonumber
\end{eqnarray}
\end{footnotesize}
where $\Phi_{2}(a,b;\rho)$ is the bivariate cumulative standard normal distribution with correlation $\rho$ evaluated at the threshold parameters $a$ and $b$. As regards the classification, in \cite{mr} it has been suggested to use an Iterative Proportional
Fitting based on the pairwise posterior probabilities obtained as output of
the pairwise EM algorithm in order to approximate the joint posterior probabilities.

\section{Model identifiability}
Model identifiability is a crucial issue, especially when latent variables are involved in conjunction with ordinal data. The necessary conditions to identify a mixture model for ordinal data using a pairwise likelihood approach are discussed in detail in \cite{mr}.  
Here we report only the necessary condition needed to identify the SCR model. We recall that the pairwise likelihood is obtained by the product of all bivariate marginal likelihood
contributions and thus the maximum number of estimable parameters is equal to the number of non redundant parameters involved in the bivariate marginals. This equals the number of parameters of a log linear model with only two factor interaction terms. As a consequence, given a $C_{1}\times C_{2}\times\ldots \times C_{P}$ contingency table a necessary condition for the identifiability of a model is that the number of model parameters is at most \begin{equation}
\sum_{i=1}^{P}(C_i-1)+\sum_{i=1}^{P-1}\sum_{j=i+1}^{P}(C_i-1)(C_j-1).
\end{equation}
Furthermore, under the URV approach, the means and the variances of the first order latent variables are fixed to 0 and 1, respectively, because they are not identified. In \cite{mr}, the authors set the means and the variances of the reference component to 0 and 1, respectively. This identification constraint individualizes uniquely the mixture components (ignoring the label switching problem), as well described in \cite{millsap04}. This is sufficient to estimate both thresholds and component parameters if all the observed variables have three categories at least and when groups are known. 
As described in the following, given the particular structure of the mean vectors and covariance matrices, it is preferable to adopt an alternative (but equivalent) parametrization. This is analogous to that one used by \cite{jor96}; it consists in setting the first two thresholds to 0 and 1, respectively. This means that there is a one-to-one correspondence between the two sets of parameters.\\
Some other parameters in the covariance matrices can be set to a specified value without loss of generality. To see this, let us consider a generic configuration for the model parameters $\bfpsi$.
$\textbf{A}$ is a non-singular $P \times P$ matrix; this can be decomposed into two sub-matrices $\textbf{A}=\left[\textbf{A}_1,\textbf{A}_2\right] $ such that the covariance matrix $\bfSigma_1$ can be written as
\begin{eqnarray*}
\bfSigma_{1}&=&\textbf{A}\begin{bmatrix}
\bfOmega_{1} & \textbf{0}\\
\textbf{0} & \bfOmega_0
\end{bmatrix} \textbf{A}^{\prime}\\
&=&\textbf{A}\begin{bmatrix}
\bfOmega_{1} & \textbf{0}\\
\textbf{0}& \textbf{0}
\end{bmatrix} \textbf{A}^{\prime}
+\textbf{A}\begin{bmatrix}
\textbf{0} & \textbf{0}\\
\textbf{0} & \bfOmega_0
\end{bmatrix} \textbf{A}^{\prime}\\
&=&\textbf{A}_1\bfOmega_{1}\textbf{A}^{\prime}_1+\textbf{A}_2\bfOmega_0\textbf{A}^{\prime}_2.
\end{eqnarray*}
In factor analysis, it is well known that there exist non-singular matrices $\textbf{S}_1$ and  $\textbf{S}_2$ such that $\textbf{A}_1\bfOmega_1\textbf{A}_1^{\prime}=\textbf{V}_1\textbf{V}_1^{\prime}$ and $\textbf{A}_2\bfOmega_0\textbf{A}_2^{\prime}=\textbf{V}_2\textbf{V}_2^{\prime}$, where $\textbf{A}_1=\textbf{V}_1\textbf{S}_1$ and $\textbf{A}_2=\textbf{V}_2\textbf{S}_2$. The matrices $\textbf{V}_1$ and $\textbf{V}_2$ have a particular structure. Assuming that $P=5$ and $Q=3$, $\textbf{V}_1$ is of order $5 \times 3$ and it looks like
\begin{eqnarray*}
\textbf{V}_1=\begin{bmatrix}
v_{11} & 0 & 0\\
v_{21}& v_{22} & 0\\
v_{31}& v_{32} &v_{33}\\
 v_{41}& v_{42}& v_{43}\\
  v_{51}& v_{52}& v_{53}\\
  \end{bmatrix};
\end{eqnarray*}
while  $\textbf{V}_2$ is of order $5 \times 2$ and it looks like
\begin{eqnarray*}
\textbf{V}_2=\begin{bmatrix}
v_{11} & 0 \\
v_{21}& v_{22} \\
v_{31}& v_{32} \\
 v_{41}& v_{42}\\
  v_{51}& v_{52}\\
  \end{bmatrix}.
\end{eqnarray*}
In other words, $\textbf{V}_1$  and $\textbf{V}_2$ have a lower triangular matrix in the first $Q$ and $(P-Q)$ rows, respectively.\\
As regards $\bfSigma_g$ with $g=2,\ldots,G$ it follows that
\begin{eqnarray*}
\bfSigma_g&=&\textbf{A}_1\bfOmega_g\textbf{A}_1^{\prime}+\textbf{A}_2\bfOmega_0\textbf{A}_2\\
&=&\textbf{V}_1\textbf{S}_1\bfOmega_g\textbf{S}_1^{\prime}\textbf{V}_1^{\prime}+\textbf{V}_2\textbf{V}_2^{\prime}\\
&=& \textbf{V}_1\bfOmega_g^{\star}\textbf{V}_1^{\prime}+\textbf{V}_2\textbf{V}_2^{\prime}.
\end{eqnarray*}
Finally, the factorization shown above does not create any problem on the structure of the mean vectors. Indeed, we observe that
\begin{eqnarray*}
\bfmu_g&=&\begin{bmatrix} \textbf{A}_1 & \textbf{A}_2
\end{bmatrix}\begin{bmatrix}\bfeta_g\\\bfeta_0\end{bmatrix}=\begin{bmatrix}\textbf{V}_1\textbf{S}_1 &\textbf{V}_2\textbf{S}_2
\end{bmatrix}\begin{bmatrix}\bfeta_g\\\bfeta_0\end{bmatrix}\\
&=&\begin{bmatrix}\textbf{V}_1 &\textbf{V}_2
\end{bmatrix}\begin{bmatrix}\textbf{S}_1\bfeta_g\\\textbf{S}_2\bfeta_0\end{bmatrix}=\begin{bmatrix}\textbf{V}_1 &\textbf{V}_2
\end{bmatrix}\begin{bmatrix}\bfeta^{\star}_g\\\bfeta_0^{\star}\end{bmatrix},
\end{eqnarray*}
where $\textbf{S}_2$ is a matrix of order $(P-Q)\times (P-Q)$. Thus the number of parameters needed to estimate the model with $Q$ variables carrying classification power, $\bar{Q}$ noisy variables and $G$ components is given by
\begin{center}
\begin{tiny}
$ 
 \underbrace{G-1}_{\mbox{$p_1,\ldots, p_{G-1}$}}+\underbrace{Q(Q+1)/2+Q(P-Q)}_{\mbox{$\textbf{V}_1$}}+\underbrace{(G-1)Q(Q+1)/2}_{\mbox{$\bfOmega_2^{\star},\ldots,\bfOmega_G^{\star}$}}+     \underbrace{(P-Q)(P-Q+1)/2+Q(P-Q)}_{\mbox{$\textbf{V}_2$}}
+\underbrace{GQ}_{\mbox{$\bfeta_1^{\star},\ldots,\bfeta_G^{\star}$}}+\underbrace{P-Q}_{\mbox{$\bfeta_{0}$}}+
\underbrace{\sum_{i=1}^{P}C_{i}-3P.}_{\mbox{$thresholds$}}
$
\end{tiny}
\end{center}
This should be less or equal to the maximum number needed to saturate a log linear model with two factor interaction terms in (5).
\section{Interpretation of matrix \textbf{A}}
As said previously, the ordinal variables are assumed to  be a partial manifestation of first-order latent variables, which are a linear combinations of second-order latent variables. The main role of matrix $\textbf{A}$ is to specify the coefficients of these linear combinations and thus, to identify the noisy variables. However this arises some interpretation issues, as it occurs in a factor analysis framework.
The solutions provided in literature are different, but they share the same idea: yielding a sparse (simple)  matrix to have an easier interpretation. To this aim, varimax and oblimin are the most popular types of rotation frequently used in the orthogonal and non-orthogonal cases, respectively. There exist many ways for creative thinking on a easier interpretation. For the current proposal we could apply a varimax rotation on $\textbf{A}_2$ and $\textbf{A}_1$. \\
Furthermore, the matrix $\textbf{A}$ plays a central role in estimating the correlation between the latent variables of first and second orders, whose covariance matrix is given by $Cov(\textbf{Y},\tilde{\textbf{Y}})=\textbf{A}\bfSigma_{\tilde{\textbf{y}}}$;
we remark that $Cov(\tilde{\textbf{Y}})$ accounts for both the within and the between variance of the mixture.
The observed variables that are most correlated with variables $\tilde{\textbf{y}}^{\bar{Q}}$ are identified as noise variables.
\\

\section{Model Selection}
In the estimation procedure, we assume that both the number of 
mixture components and the number of noisy variables are fixed. In practice, they are unknown and
thus, they must be estimated through observed data
The best fitted model is chosen by selecting the model minimizing the C-BIC, introduced by \cite{gao2010}.
\begin{equation}
\mbox{C-BIC}=-2p\ell(\hat{\bfpsi};\textbf{x}) + \mbox{tr}\left(\hat{\bf{H}}^{-1}\hat{\bf{V}} \right)\log N.
\end{equation}
where $\textbf{H}$ is the sensitivity matrix, $\textbf{H}=E(-\nabla ^{2}p\ell (\bfpsi;\textbf{x}))$ while $\textbf{V}$ is the variability matrix (the covariance matrix of the score vector), $\textbf{V}=Var(\nabla p\ell (\bfpsi;\textbf{x}))$.
The C-BIC has the same structure of BIC; the only difference is the way used to account for the model complexity. The BIC penalizes the likelihood by the term $d\log N$, where $d$ is the total number of essential parameters. On the other hand C-BIC penalizes the likelihood by $\mbox{tr}\left(\hat{\bf{H}}^{-1}\hat{\bf{V}} \right)\log N$. In this case, the identity $\textbf{H}=-\textbf{V}$ does not hold, since the likelihood components are not independent (differently from the full likelihood theory). However, if  $\textbf{H}=-\textbf{V}$, then $\mbox{tr}\left(\hat{\bf{H}}^{-1}\hat{\bf{V}} \right)$ would be equal to $d$. 
Sample estimates of $\textbf{H}$ and $\textbf{V}$ for the model proposed are 
\begin{center}
$\hat{\bf{H}}=-\frac{1}{N}\sum_{r=1}^{R}n_r\nabla ^{2}p\ell (\hat{\bfpsi};\textbf{x})$
\end{center}
and
\begin{center}
$\hat{\bf{V}}=\frac{1}{N}\sum_{r=1}^{R}n_r(\nabla p\ell (\hat{\bfpsi};\textbf{x}))(\nabla p\ell (\hat{\bfpsi};\textbf{x}))^{\prime}.$
\end{center}
A simulation study testing its performance in a context of mixture models has been provided in \cite{mr,mr1}. 
In the current work, in order to obtain the empirical estimates of the sensitivity and variability matrices, we have used the same numerical approximation technique described there. More precisely, the derivatives are estimated by finite differences. As regards the variability matrix a covariance matrix of the score function has been estimated for each response pattern. Computationally speaking it has been obtained by multiplying a matrix including the score functions for each response pattern times a diagonal matrix with the frequencies $n_r$ on the main diagonal times the first matrix transposed. As regards the sensitivity matrix, we know from the theoretical results of the pairwise that each sub-likelihood (each component of the pairwise likelihood) is a true likelihood. This means that the second Bartlett's identity holds. This allows us to estimate the sensitivity matrix in the same fashion as before. However in this case the diagonal matrix has the frequencies $n_{x_{i}x_{j}}$ on the main diagonal and the score functions refer to each response pattern for each pair of variables. Finally, the trace is obtained by summing the generalized eigenvalues of the two matrices, i.e. by solving the equation $\hat{\textbf{V}}\textbf{x}=\lambda\hat{\textbf{H}}\textbf{x}$. This allows to avoid inverting the sensitivity matrix, that may be imprecise and unstable.

\section{Simulation study}
To evaluate the empirical behaviour of the proposal, a large-scale simulation study has been conducted. The performance has been evaluated in terms of recovering the true cluster structure using the following measures:
the loss measure (L) between the estimated and the true model and
the Adjusted Rand Index (ARI) \cite{hubert85} between
the true hard partition matrix and that estimated. The former compares clusterings by set matching and it is given by the quadratic mean of the differences between the true and the estimated posterior probabilities. Since, label-switching plays an important role, we compute it for every possible permutation of the cluster membership labels of the resulting partition of $N$ individuals and we choose the minimum value obtained. A smaller value clearly indicates a better performance with $0\leq L \leq 1$. The second index can be considered a hard classification measure, while the former a fuzzy index. 
Given two different hard classification matrices,  $\textbf{W}$ and $\hat{\textbf{W}}$,  i.e. binary row  matrices according to which observations are assigned to only one cluster, the ARI counts the pairs of observations that are assigned to the same or different clusters under both partition matrices and it is defined as
\begin{equation}
ARI(\textbf{W},\hat{\textbf{W}})=\dfrac{R(\textbf{W},\hat{\textbf{W}})-E(R(\textbf{W},\hat{\textbf{W}}))}{1-E(R(\textbf{W},\hat{\textbf{W}}))},
\end{equation}
where 
\begin{equation*}
R(\textbf{W},\hat{\textbf{W}})=\dfrac{N_{11}+N_{00}}{\binom {N} {2} },
\end{equation*}
where $R(\textbf{W},\hat{\textbf{W}})$ is the Rand Index, $\textbf{W}$ and $\hat{\textbf{W}}$ are the true and the estimated partition matrices respectively, $N_{11}$ is the number of pairs of observations in the same cluster under $\textbf{W}$ and $\hat{\textbf{W}}$ and $N_{00}$ is the number of pairs in different clusters under $\textbf{W}$ and $\hat{\textbf{W}}$; $N$ is the sample size. 
The index has expected value zero for independent clusterings and maximum value 1 for identical clusterings.\\
Eight different scenarios have been considered under the presence or not of noise variables. In both cases we simulated 250 samples from a latent two-component Gaussian mixture model. However, in the first case 
we simulated five ordinal variables with five categories, but we assumed that only two ($Q=2$) variables carry  group discrimination information, the others are noise variables. In the second case, we assumed that three variables are less informative about cluster structure. Nevertheless, their means and variances still change across the groups (differently from the assumptions of the SCR model).\\
Under these two broad conditions, we have analysed four scenarios considering two different experimental factors: the sample size ($N=1000,5000$) and the separation between clusters (well separated or not).
\\
Given the simulated ordinal data, we compared the performances of the SCR model with the \textit{standard} clustering model proposed in \cite{mr}. The parameter estimates
were carried out through a pairwise EM algorithm, that has been initialized using \textit{rational} starting points. In other words,  we first fitted a Gaussian mixture model, treating the ranks as continuous. Then, we used its output properly. The algorithms were stopped when the increase in the asymptotic estimate log-likelihood between two consecutive steps was less than $10^{-2}$.\\
In the sequel, we analyze the simulation output  in the case in which three noise variables exist; then, we analyze the case in which three variables are less informative about the cluster structure. This section ends with a comparison between these two main conditions.\\
Below, we report the true values used to generate the data according to the SCR model, i.e. the case in which there are three noise variables.
\begin{table}[!htbp]
\begin{center}
\begin{tiny}
\caption{True values of the latent mixture model and thresholds under different scenarios. The data were generated according the structure assumed by the SCR model.}
\begin{tabular}{ccc}
\hline
\multicolumn{3}{c}{\textbf{{\footnotesize Common parameters  in terms of $\textbf{A}$, $\bfeta$ and $\bfOmega$} }} \\
\cline{1-3}\\ 
\textit{Component weights}& $p_1=0.3$ &$p_2=0.7$ \\
\textit{Means of noisy variables}&\multicolumn{2}{c}{$\bfeta_0=[0,0,0]^{\prime}$}\\
\textit{Covariance matrices}& $\bfOmega_1=\begin{bmatrix}
1& 0 \\
0 &1 \\
 \end{bmatrix}$ &$\bfOmega_0=\begin{bmatrix}
1& 0& 0\\
0& 1& 0\\
0 &0& 1\\
 \end{bmatrix}$ \\
 \textit{Thresholds} & \multicolumn{2}{c}{for each variable:$[0,1,2,3]$}\\
\hline
\hline
\multicolumn{3}{c}{\textbf{{\footnotesize Separated groups} }} \\
\hline 
 \multicolumn{3}{c}{Parametrization in terms of $\textbf{A}$, $\bfeta$ and $\bfOmega$} \\
\cline{1-3}\\ 
&$\bfeta_1=[-2.24,4.47]^{\prime}$&$
 \bfeta_2=[-2.80,0.56]^{\prime}$ \\
&  $\bfOmega_2=\begin{bmatrix}
1.25& 0.75 \\
0.75& 1.25 \\
 \end{bmatrix}$ 
 & $\textbf{A}=\begin{bmatrix}
\sqrt{.8}& 0& 0 &0 &0 \\
0& \sqrt{.8}& 0 &0 &0\\
0 &0 &\sqrt{1.5}&0 &0\\
0 &0 &0&\sqrt{1.5}&0\\
0 &0 &0& 0&\sqrt{1.5}\\
 \end{bmatrix}$ \\
 \\
   \hline
       \multicolumn{3}{c}{Parametrization in terms of  $\bfmu$ and $\bfSigma$} \\
 \hline\\ 
$\bfmu$ &$[ -2,4,0,0,0]^{\prime}$ &$[ 2.5, 0.5,0,0,0]^{\prime}$\\
$\bfSigma$&
$\begin{bmatrix}
  0.8  & 0 &        0  &       0&         0\\
   0&    0.8    &     0    &     0&         0\\
   0&        0 &   1.5  &      0   &      0\\
      0 &        0   &      0 &   1.5&         0\\
      0 &        0  &       0 &        0 &   1.5\\
\end{bmatrix}$&$
\begin{bmatrix}
1.0&   0.6 &       0   &      0&         0\\
    0.6  & 1.0   &     0     &    0      &   0\\
         0 &        0 &   1.5 &   0  &       0\\
         0       &  0  &       0  &  1.5 & 0\\
         0    &     0   &      0 &        0   & 1.5\\
\end{bmatrix}$\\
\\
 \hline
\hline
\multicolumn{3}{c}{\textbf{{\footnotesize Non-separated groups} }} \\
\hline 
 \multicolumn{3}{c}{Parametrization in terms of $\textbf{A}$, $\bfeta$ and $\bfOmega$} \\
\cline{1-3}\\ 
 &$\bfeta_1=[-0.,403,2.86]^{\prime}$&
    $\bfeta_2=[2.04,0.41]^{\prime}$ \\
&$\bfOmega_2=\begin{bmatrix}
2.3& 1.3 \\
1.3& 2.2 \\
 \end{bmatrix}$ & $\textbf{A}=\begin{bmatrix}
\sqrt{1.5}& 0& 0 &0 &0 \\
0& \sqrt{1.5}& 0 &0 &0\\
0 &0 &\sqrt{1.5}&0 &0\\
0 &0 &0&\sqrt{1.5}&0\\
0 &0 &0& 0&\sqrt{1.5}\\
 \end{bmatrix}$ \\
 \\
   \hline
       \multicolumn{3}{c}{Parametrization in terms of  $\bfmu$ and $\bfSigma$} \\
 \hline\\ 
$\bfmu$ &$[ -0.5,3.5,0,0,0]^{\prime}$ &$[ 2.5, 0.5,0,0,0]^{\prime}$\\
$\bfSigma$&
$\begin{bmatrix}
  1.5  & 0 &        0  &       0&         0\\
   0&    1.5    &     0    &     0&         0\\
   0&        0 &   1.5  &      0   &      0\\
      0 &        0   &      0 &   1.5&         0\\
      0 &        0  &       0 &        0 &   1.5\\
\end{bmatrix}$&$
\begin{bmatrix}
3.30&   1.95 &       0   &      0&         0\\
    1.95  & 3.30   &     0     &    0      &   0\\
         0 &        0 &   1.5 &   0  &       0\\
         0       &  0  &       0  &  1.5 & 0\\
         0    &     0   &      0 &        0   & 1.5\\
\end{bmatrix}$\\
\\
 \hline
\end{tabular}
\end{tiny}
\end{center}
\end{table}
\newpage

\begin{figure}[!htbp]
  \centering
  \caption{Box-plots of ARI for the posterior probabilities. Data generated
from a two-component  latent mixture; 5 ordinal variables with 5 categories; three of them are noisy variables. N=1000,5000. Separated/non-separated  groups. 250 samples.}
 \includegraphics[scale=0.3]{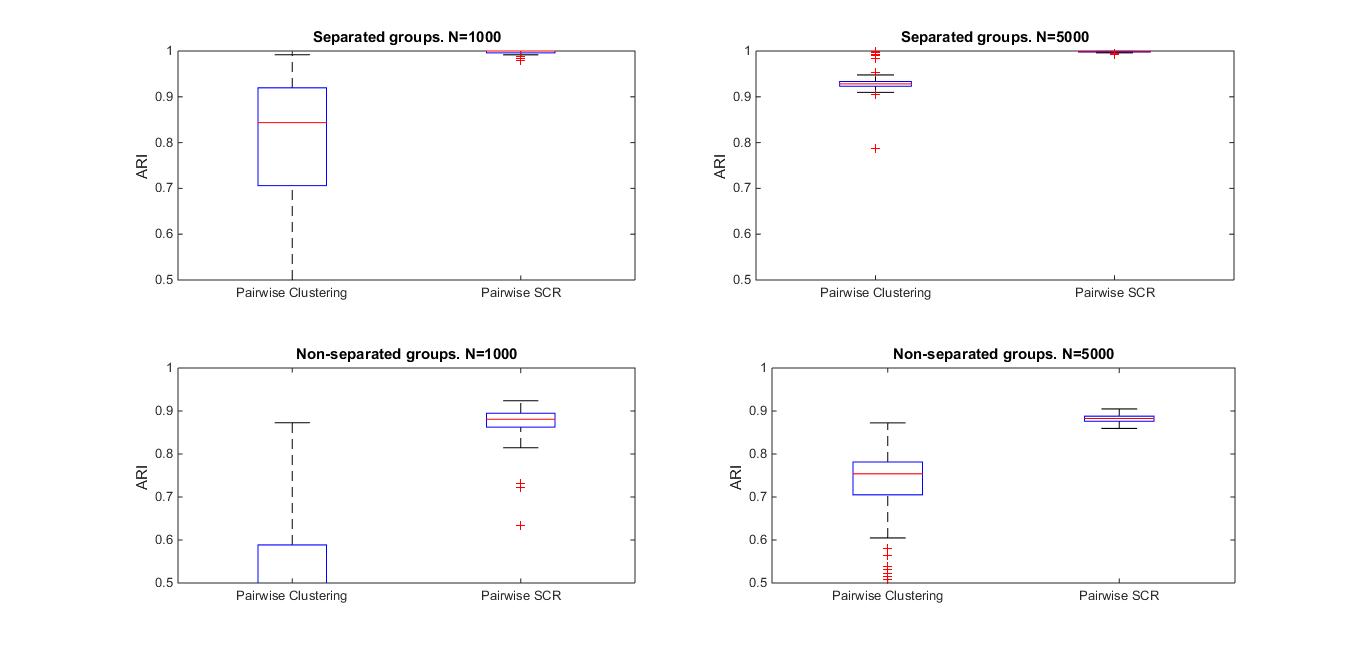}
\end{figure}
\begin{figure}[!htbp]
  \centering
  \caption{Box-plots of LOSS for the posterior probabilities. Data generated
from a two-component  latent mixture; 5 ordinal variables with 5 categories; three of them are noisy variables. N=1000,5000. Separated/non-separated  groups. 250 samples.}
  \includegraphics[scale=0.3]{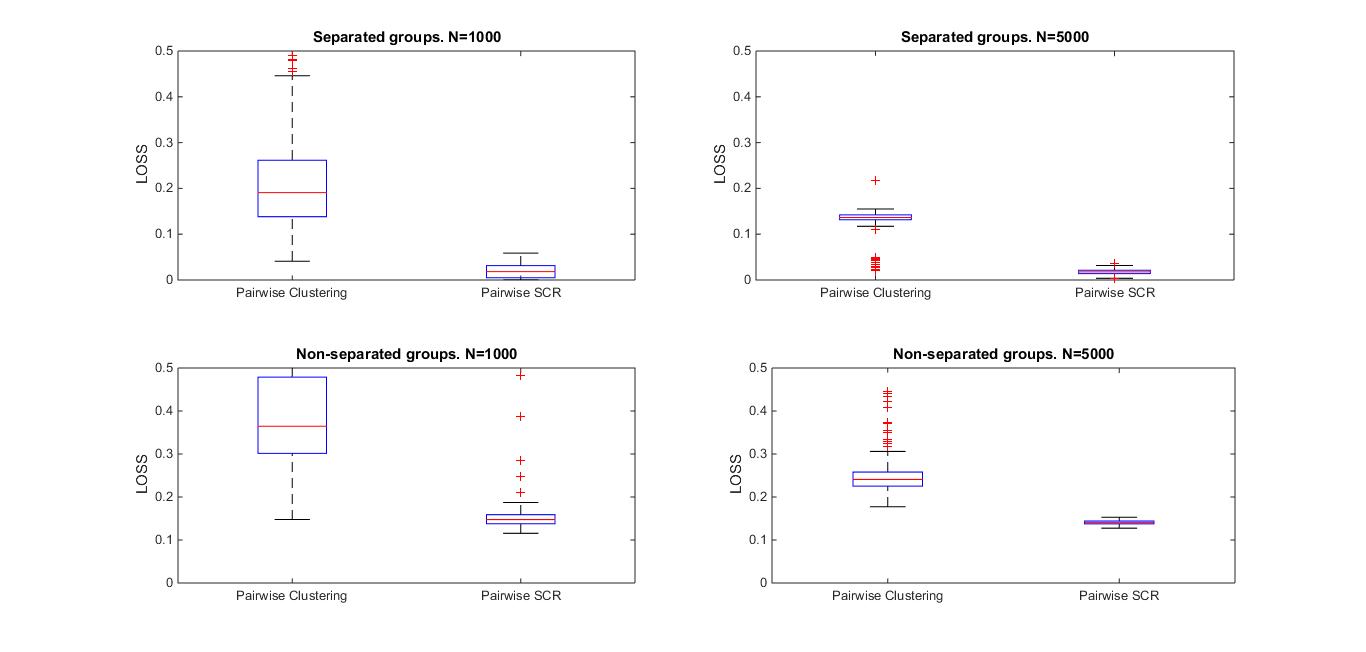}
\end{figure}
All simulation results are reported in the appendix. Figures 1 and 2 show the distributions of the adjusted rand index
and loss measure, respectively, in the four different scenarios. 
On the left side the sample size is equal to 1000,
while on the right one is equal to 5000; in the first row the
groups are separated, while in the second one the groups are not separated. To be more
clear and to have more comparable results, the range of the y-axis has been cut ([0.5, 1] and [0, 0.5] for the adjusted rand index and loss measure, respectively). 
The pairwise estimators shows consistency: as $N$ increases we obtain better classification performance and the variances of ARI and loss are smaller. Furthermore, the clustering performance becomes poorer as the components are less separated. Comparing the two fitted models, we observe that SRC outperforms the pairwise clustering in all scenarios, as expected. However, the gap in performance depends on the specific scenario. In general, the gap seems to increase when the groups are less separated and the sample size is smaller. 
\\
Now, we report the simulation results 
for the case in which the data were not generated according the structure assumed by the SCR model. They were assumed to be a categorization of a latent two-component Gaussian mixture model, whose true parameters are reported below.
There are three less informative variables; they are less informative in the sense that their means and variances change slightly over the components. In other word, based on these variables, the two components are almost totally overlapped.   

\begin{table}[!htbp]
\begin{center}
\begin{scriptsize}
\caption{True values of the latent mixture model and thresholds under different scenarios. The data were generated by thresholding a latent two-component Gaussian mixture model.}
\begin{tabular}{ccc}
\hline
\multicolumn{3}{c}{\textbf{{\footnotesize Separated groups} }} \\
& $p_1=0.3$ &$p_2=0.7$ \\
\cline{1-3}\\ 
$\bfmu$&$[ -2,4,0,-0.5,0]^{\prime}$ &$[ 2.5, 0.5,0.5,0,0.5]^{\prime}$\\
$\bfSigma$&
$\begin{bmatrix}
  0.8  & 0 &        0  &       0&         0\\
   0&    0.8    &     0    &     0&         0\\
   0&        0 &   1.5  &      0   &      0\\
      0 &        0   &      0 &   1.5&         0\\
      0 &        0  &       0 &        0 &   1.5\\
\end{bmatrix}$&$
\begin{bmatrix}
1.25&   0.75 &       0   &      0&         0\\
    0.75  & 1.25   &     0     &    0      &   0\\
         0 &        0 &   1.0 &   0  &       0\\
         0       &  0  &       0  &  1.0 & 0\\
         0    &     0   &      0 &        0   & 1.0\\
\end{bmatrix}$\\
\textit{Thresholds} & \multicolumn{2}{c}{for each variable:$[0,1,2,3]$}\\
 \hline

 \multicolumn{3}{c}{\textbf{{\footnotesize Non-separated groups} }} \\
& $p_1=0.3$ &$p_2=0.7$ \\ 
\cline{1-3}\\ 
$\bfmu$&$[ -0.5,   3.5,0,-0.5,0]^{\prime}$ &$[   2.5,0.5,0.5,0,0.5]^{\prime}$\\
$\bfSigma$&$
\begin{bmatrix}
  1.5  & 0 &        0  &       0&         0\\
   0&    1.5    &     0    &     0&         0\\
   0&        0 &   1.5  &      0   &      0\\
      0 &        0   &      0 &   1.5&         0\\
      0 &        0  &       0 &        0 &   1.5\\
\end{bmatrix}$&$
\begin{bmatrix}
2.2&   1.3 &       0   &      0&         0\\
    1.3  & 2.2   &     0     &    0      &   0\\
         0 &        0 &   1.0 &   0  &       0\\
         0       &  0  &       0  &  1.0 & 0\\
         0    &     0   &      0 &        0   & 1.0\\
\end{bmatrix}$\\
\textit{Thresholds} & \multicolumn{2}{c}{for each variable:$[0,1,2,3]$}\\
 \hline
\end{tabular}
\end{scriptsize}
\end{center}
\end{table}
\newpage
\begin{figure}[!htbp]
  \centering
  \caption{Box-plots of ARI for the posterior probabilities. Data generated
from a latent two-component mixture model; 5 ordinal variables with 5 categories; three of them are less informative about the cluster structure. N=1000,5000. Separated/non-separated  groups. 250 samples.}
\includegraphics[scale=0.3]{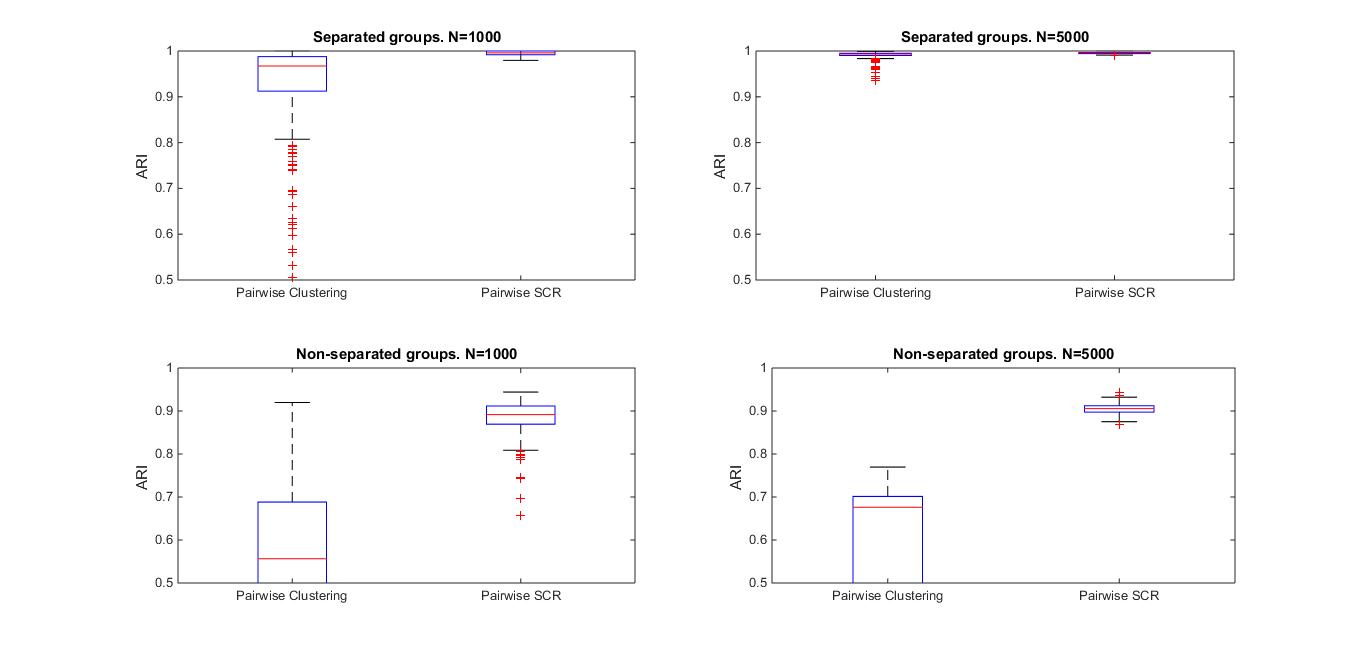}
\end{figure}

\begin{figure}[!htbp]
  \centering
  \caption{Box-plots of LOSS for the posterior probabilities. Data generated
from a latent two-component mixture model; 5 ordinal variables with 5 categories; three of them are less informative about the cluster structure. N=1000,5000. Separated/non-separated groups. 250 samples.}
\includegraphics[scale=0.3]{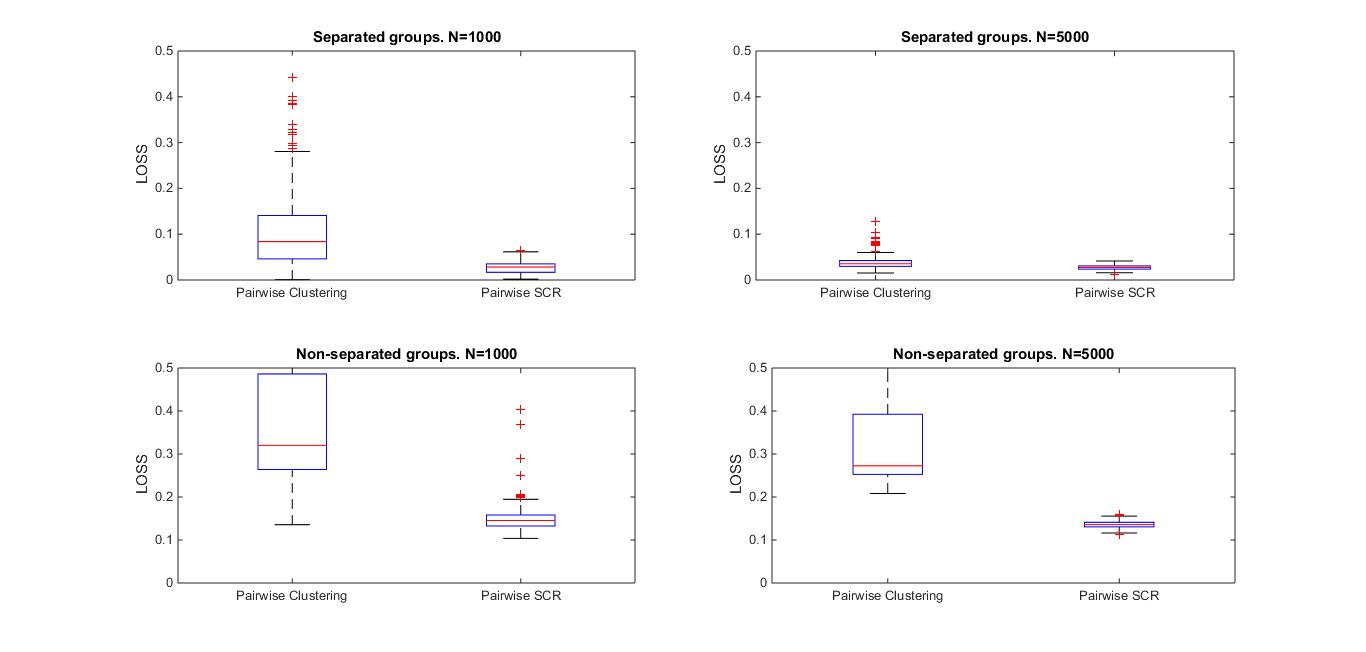}
\end{figure}
All simulation results are reported in the appendix. Figures 3 and 4 show the distributions of the adjusted rand index
and loss measure, respectively, in the four different scenarios. Once again, the pairwise estimators shows consistency. As the degree of overlap between components increases, the performances worsen. Comparing the two fitted models, the only scenario in which their performances are almost the same is the \textit{easiest}, i.e. when the groups are separated and the thresholds are equidistant. In all other scenarios, it seems that the presence of three less informative variables mask the cluster structure, and therefore this is not successfully recovered by the pairwise clustering model. Conversely, the SCR model recognizes the presence of some noise dimensions and identifies the two variables carrying the discriminative classification information, using less parameters (in other words it a more parsimonious model). This leads to better results in terms of clustering performances. 
\\
Finally, we compare briefly the two main conditions: the existence of noise variables versus the existence of less informative variables. When there are less informative variables (i.e. looking at Figures 3 and 4) we note that the performances of pairwise clustering model improve, compared to Figures 1 and 2. This is somehow expected, since in the last case, even if the cluster structure could be masked, there is no mis-specification between the generating data process and the fitting model. Nevertheless, due to the presence of some less informative variables makes its performances, it is still outperformed by the SCR model. On the other hand, this shows some degree of robustness for the SCR model; in other words, even if the data were generated from a mis-specified model, this does no effect its performances.


\section{Application to Real Data}
In this section the proposed modelling methodology is applied to a real dataset. \\

\subsection{General Social Survey dataset}
To illustrate how the model can be used we apply it to a set of data taken from the General Social Survey and displayed in Table 3. This is a well known dataset in educational field, analysed by \cite{goodman84} and re-analysed recently by \cite{giordan11} and \cite{mr}. It is a three-way cross-classification table of 1,517 people
on three ordinal variables: happiness (3 categories), years of completed schooling
(4 categories), and number of siblings (5 categories). 

\begin{table}[!htbp]
\begin{center}
\begin{scriptsize}
\caption{ Three-way cross-classification of U.S. sample according to their reported happiness, years of schooling and number of siblings}
\begin{tabular}{cccccc}
\hline
& \multicolumn{5}{c}{\textit{Number of Siblings}}\\
\textit{Year of School} \cr
\cline{2-6}
\textit{Completed} &0-1& 2-3& 4-5&6-7&8+\\
\hline
& \multicolumn{5}{c}{\textit{Not too Happy}}\\
\cline{2-6}
$<12$& 15&34&36&22&61\\
12&31&60&46&25&26\\
13-16&35&45&30&13&8\\
17+&18&14&3&3&4\\
\cline{2-6}
& \multicolumn{5}{c}{\textit{Pretty Happy}}\\
\cline{2-6}
$<12$& 17&53&70&67&79\\
12&60&96&45&40&31\\
13-16&63&74&39&24&7\\
17+&15&15&9&2&1\\
\cline{2-6}
& \multicolumn{5}{c}{\textit{Very Happy}}\\
\cline{2-6}
$<12$& 7&20&23&16&36\\
12&5&12&11&12&7\\
13-16&5&10&4&4&3\\
17+&1&2&9&0&1\\
\hline
\end{tabular}
\end{scriptsize}
\end{center}
\end{table}
We initialized the pairwise EM algorithm considering 100 different random starting points. We run 9 different scenarios varying both the number of clusters $G=1,2,3$ and the number of variables with classification power $Q=1,2,3$.
All models with \textit{G} greater than 3 cannot be identified.
The final model is chosen by minimizing the C-BIC. \\
The best fitted model is given by $G=2$ and $Q=1$ (see Table 2), with the component weights equal to 0.28 and 0.72, respectively. Figure 3 represents the posterior probabilities to belong to the largest component. It is worth noting that there is a clear classification between the two groups as the number of completed years of schooling increases. Moreover it is interesting to note that years of completed schooling is the only variable with discriminative power, since the posterior probabilities do not change substantially over the levels of happiness or the number of siblings.
\begin{figure}
       \centering
         \caption{Heat map of posterior probabilities}
\includegraphics[scale=0.15]{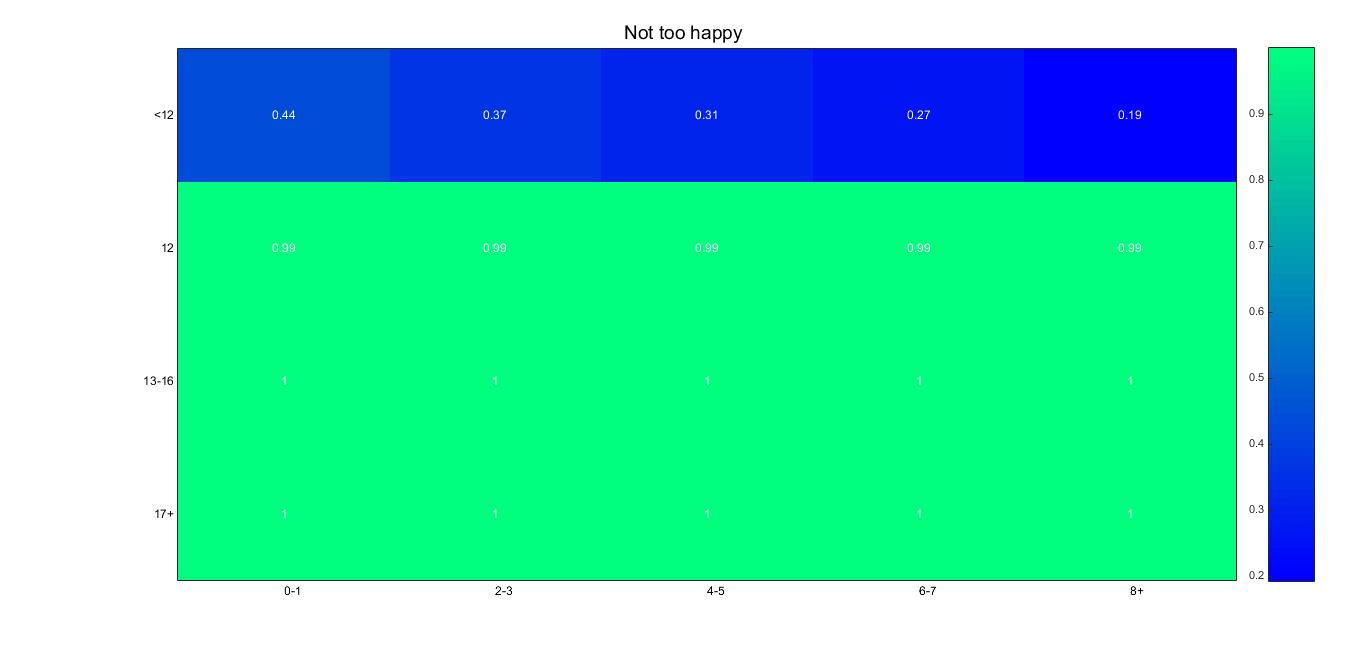}                \includegraphics[scale=0.15]{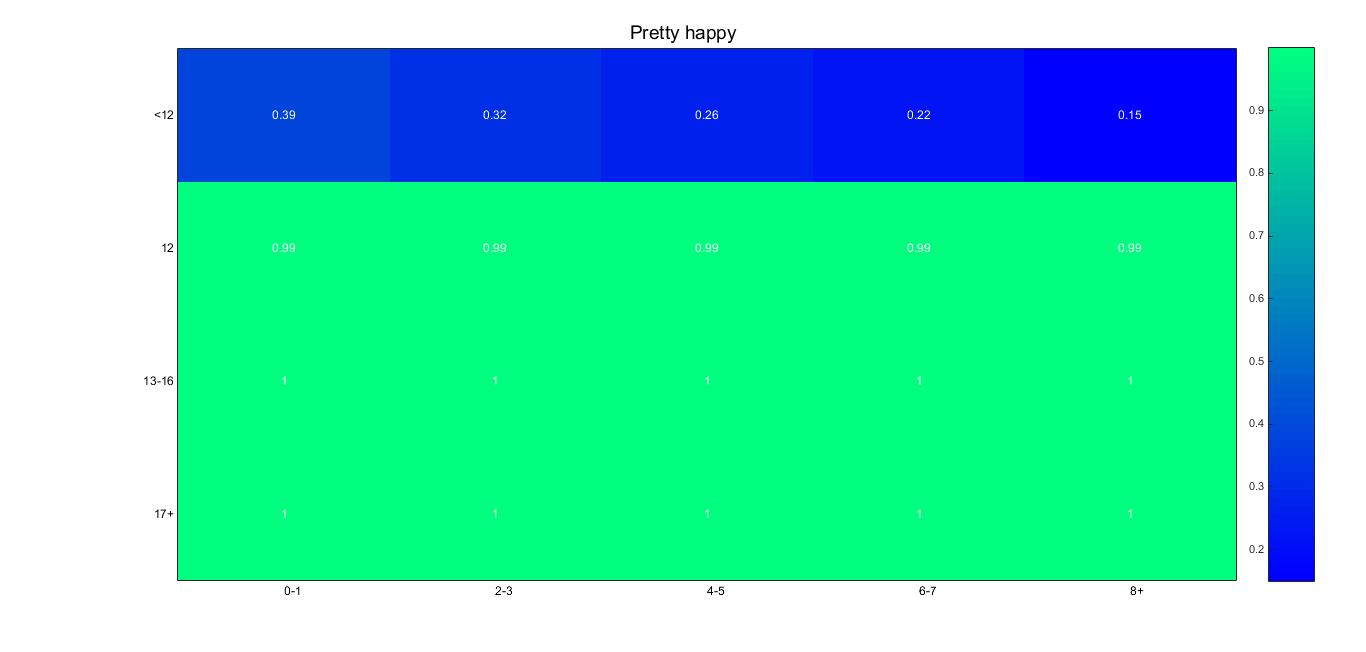}
\includegraphics[scale=0.15]{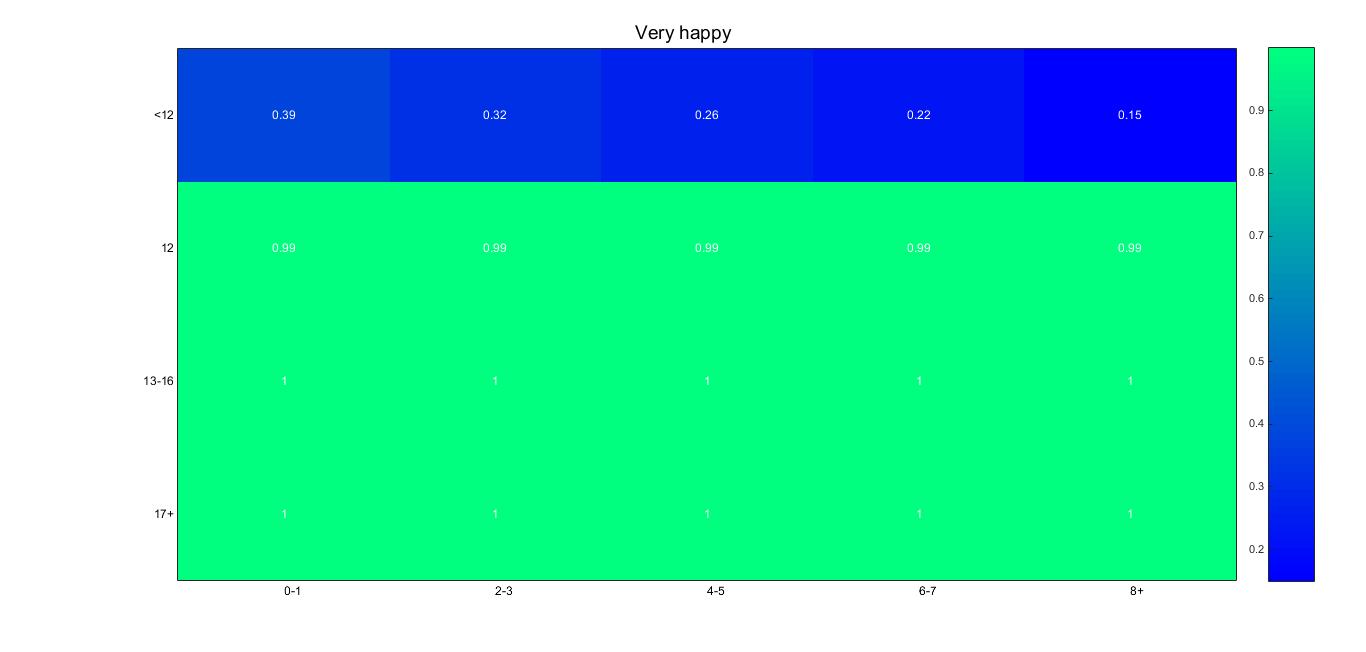}
                              \end{figure}
%
\begin{table}[!htbp]
\begin{center}
\begin{scriptsize}
\caption{ Model choice according to the composite information criteria  C-BIC.}
\begin{tabular}{cccc}
\hline 
 &\textbf{ G=1 }& \textbf{G=2}&\textbf{G=3}\\
\hline
Q=1&24717  & \textbf{22848}&22890 \\ 
 \hline
 Q=2& 23151  &  22881  &22891   \\ 
  \hline
   Q=3&22937& 22896& 22972\\
 \hline
\end{tabular}
\end{scriptsize}
\end{center}
\end{table}

The correlation between the first- and second-order variables (by rows and by columns, respectively) is
\begin{center}
$\left[\,
\begin{array}{  c : c c } 
0.9997 &  -0.0797  & -0.0094\\
\hdashline
      -0.4763  &  0.8977  & -0.0988\\
      -0.1740  &  0.1015   & 0.9824\\
\end{array}\,\right]$
\end{center}

This leads to some straightforward conclusions: to detect the noisy variables we should look at the highest correlation on the last two columns ($\tilde{y}_2, \tilde{y}_3$). The most correlated variables are $y_2$ and $y_3$ with correlations equal to 0.90 and 0.98, respectively.\\
Furthermore, in order to test the right behaviour of our proposal, in the original dataset we have included a noisy ordinal variable with three categories obtained by thresholding a standard normal variable.
%
%
%
\begin{table}[!htbp]
\begin{center}
\begin{scriptsize}
\caption{ Model choice according to the composite information criteria C-BIC.}
\begin{tabular}{cccc}
\hline 
 &\textbf{ G=1 }& \textbf{G=2}&\textbf{G=3}\\
\hline
 Q=1 & 44407   & \textbf{44133}  &  44151   \\
  \hline
  Q=2 & 44719  & 44182 & 44166     \\
  \hline
 Q=3 &  44423  & 44162   & 44186     \\
 \hline
Q=4 & 44809  & 44219  & 44313      \\
\hline
\end{tabular}
\end{scriptsize}
\end{center}
\end{table} 
As expected the best fitted model is that one minimizing C-BIC, that is the model with $G=2$ and $Q=1$ with a C-BIC value of 44133.
\\
The correlation between the first- and second-order variables (by rows and by columns, respectively) is
\begin{center}
$\left[\,
\begin{array}{  c : c c c}
0.9986  & -0.1726  & -0.0259 &  -0.0096\\\hdashline
     -0.4800  &  0.9286  &  0.0286  &  0.0019\\
    -0.2157 &   0.0349  &  0.9816  &  0.0038\\
     -0.0580 &  -0.0119 &   0.0260 &   0.9985\\
\end{array}\,\right]$
\end{center}
This leads to some straightforward conclusions: to detect the noisy variable we should look at the highest correlation on the last three columns ($\tilde{y}_2,\tilde{y}_3,\tilde{y}_4$). The most correlated variables are $y_2,y_3,y_4$ with correlations equal to $0.93,0.98$ and $1$, respectively.\\

\section{Concluding Remarks}

In this paper an extension of the model proposed by \cite{everitt88, lubke08,mr} has been introduced. The proposal allows to select the variables that are significant for clustering. Indeed in many applications, it is possible that only some variables have classification power. From a statistical modelling point of view, this means requiring a particular structure for the means and the covariance matrices. Following the URV approach the ordinal variables are considered a partial manifestation of first-order latent variables. To detect the presence of noisy variables and/or dimensions, these are assumed to be linear combinations of two independent sets of second-order latent variables. Such proposal reduces and clusters ordinal data simultaneously. Nevertheless if there is no noisy variable, but only noisy dimensions, it reduces to a more parsimonious mixture model to cluster ordinal data (compared to the proposals existing in literature).
Whatever the structure is (apart from the independence case), the full likelihood always involves multidimensional integrals that cannot be computed in a closed form. For this reason, the parameter estimation is carried out through the maximization of an easier surrogate function, that is the pairwise likelihood.  In order to classify the observations, the posterior probabilities are re-constructed through the IPF algorithm. After exploring the effectiveness of the proposal through a large-scale simulation study, an application to real dataset has been analysed. To validate the proposal, a further experiment has been conducted: an ordinal noisy variable has been added to the original General Social Survey dataset. In all cases the best fitted model has been chosen by minimizing the information criterion C-BIC.
\\
Even if the proposal seems to be promising, there are some open issues. For example, in the current work we do not provide a graphical representation of the output in a reduced space. It is not straightforward for two main reasons: ordinal variables do not have a friendly graphical representation and furthermore, there exist two different orders of latent variables. However, this \textit{challenge} gives us motivation for further research.

\newpage

\section*{Appendix}
\subsection{Data generated from the SCR model}
\begin{table}[!htbp]
\begin{center}
\begin{scriptsize}
\caption{ Simulation results: ARI and loss for the posterior probabilities. Data generated
from a two-component  latent mixture; 5 ordinal variables with 5 categories; three of them are noisy variables. Separated groups. N=1000 and R=250 samples.}
\begin{tabular}{cccccccc}
\cline{1-8}\\
& \multicolumn{7}{c}{\textbf{Adjusted Rand Index}}\\ 
&\textbf{Mean} & \textbf{St.Dev}&  \textbf{q=0.025}&\textbf{q=0.25}&\textbf{q=0.5}&\textbf{q=0.75}&\textbf{q=0.975} \\ 
\cline{2-8}\\
Pairwise \textit{C}&   0.7518 &   0.2478 &   0.5446 &   0.7543  &  \textbf{0.8436} &   0.9004    &0.9385 \\
Pairwise \textit{SCR}&    0.9970  &  0.0040 &   0.9958  &  0.9959 &  \textbf{1.0000} &   1.0000  &  1.0000 \\
\cline{1-8}\\
& \multicolumn{7}{c}{\textbf{Loss}}\\ 
&\textbf{Mean} & \textbf{St.Dev}&  \textbf{q=0.025}&\textbf{q=0.25}&\textbf{q=0.5}&\textbf{q=0.75}&\textbf{q=0.975} \\ 
\cline{2-8}\\
Pairwise \textit{C}&    0.2230 &   0.1247   & 0.1216  &  0.1560  & \textbf{ 0.1908 }&   0.2305  &  0.3016 \\
Pairwise \textit{SCR}&   0.0190 &   0.0143 &   0.0032  &  0.0082 &   \textbf{0.0182}  &  0.0283  &  0.0333 \\
\cline{1-8}
\end{tabular}
\end{scriptsize}
\end{center}
\end{table}

\begin{table}[!htbp]
\begin{center}
\begin{scriptsize}
\caption{ Simulation results: ARI and loss for the posterior probabilities. Data generated
from a two-component  latent mixture; 5 ordinal variables with 5 categories; three of them are noisy variables. Separated groups. N=5000 and R=250 samples.}
\begin{tabular}{cccccccc}
\cline{1-8}\\
& \multicolumn{7}{c}{\textbf{Adjusted Rand Index}}\\ 
&\textbf{Mean} & \textbf{St.Dev}&  \textbf{q=0.025}&\textbf{q=0.25}&\textbf{q=0.5}&\textbf{q=0.75}&\textbf{q=0.975} \\ 
\cline{2-8}\\
Pairwise \textit{C}&       0.9300  &  0.0178&    0.9204  &  0.9251  &  \textbf{0.9283}  &  0.9323  &  0.9353 \\
Pairwise \textit{SCR}&       0.9985 &   0.0013  &  0.9975  &  0.9984 &   \textbf{0.9984}  &  0.9992 &   1.0000 \\
\cline{1-8}\\
& \multicolumn{7}{c}{\textbf{Loss}}\\ 
&\textbf{Mean} & \textbf{St.Dev}&  \textbf{q=0.025}&\textbf{q=0.25}&\textbf{q=0.5}&\textbf{q=0.75}&\textbf{q=0.975} \\ 
\cline{2-8}\\
Pairwise \textit{C}&       0.1326  &  0.0229 &   0.1288  &  0.1331   & \textbf{0.1365}  &  0.1394  &  0.1439\\
Pairwise \textit{SCR}&  0.0174 &   0.0065 &   0.0109 &   0.0148   & \textbf{0.0178}   & 0.0200  &  0.0235  \\
\cline{1-8}
\end{tabular}
\end{scriptsize}
\end{center}
\end{table}

\begin{table}[!htbp]
\begin{center}
\begin{scriptsize}
\caption{ Simulation results: ARI and loss for the posterior probabilities. Data generated
from a two-component  latent mixture; 5 ordinal variables with 5 categories; three of them are noisy variables. Non-separated groups. N=1000 and R=250 samples.}
\begin{tabular}{cccccccc}
\cline{1-8}\\
& \multicolumn{7}{c}{\textbf{Adjusted Rand Index}}\\ 
&\textbf{Mean} & \textbf{St.Dev}&  \textbf{q=0.025}&\textbf{q=0.25}&\textbf{q=0.5}&\textbf{q=0.75}&\textbf{q=0.975} \\ 
\cline{2-8}\\
Pairwise \textit{C}&     
    0.3696   & 0.2519   & 0.0619  &  0.2071 &   \textbf{0.4022}  &  0.5364   & 0.6481\\
Pairwise \textit{SCR}&   0.8722  &  0.0649   & 0.8544  &  0.8685  &  \textbf{0.8809 }  & 0.8915 &   0.9002\\
\cline{1-8}\\
& \multicolumn{7}{c}{\textbf{Loss}}\\ 
&\textbf{Mean} & \textbf{St.Dev}&  \textbf{q=0.025}&\textbf{q=0.25}&\textbf{q=0.5}&\textbf{q=0.75}&\textbf{q=0.975} \\ 
\cline{2-8}\\
Pairwise \textit{C}&    0.3968  &  0.1194 &   0.2874  &  0.3216  & \textbf{ 0.3646}   & 0.4345 &   0.5237\\
Pairwise \textit{SCR}&       0.1517  &  0.0317&    0.1340 &   0.1414  &  \textbf{0.1475} &   0.1536 &    0.1655\\
\cline{1-8}
\end{tabular}
\end{scriptsize}
\end{center}
\end{table}

\begin{table}[!htbp]
\begin{center}
\begin{scriptsize}
\caption{ Simulation results: ARI and loss for the posterior probabilities. Data generated
from a two-component  latent mixture; 5 ordinal variables with 5 categories; three of them are noisy variables. Non-separated groups. N=5000 and R=250 samples.}
\begin{tabular}{cccccccc}
\cline{1-8}\\
& \multicolumn{7}{c}{\textbf{Adjusted Rand Index}}\\ 
&\textbf{Mean} & \textbf{St.Dev}&  \textbf{q=0.025}&\textbf{q=0.25}&\textbf{q=0.5}&\textbf{q=0.75}&\textbf{q=0.975} \\ 
\cline{2-8}\\
Pairwise \textit{C}&      0.7276  &  0.1034   & 0.6850  &  0.7255 &   \textbf{0.7539}   & 0.7731   & 0.7914\\
Pairwise \textit{SCR}&   0.8823  &  0.0086   & 0.8736 &   0.8789 &   \textbf{0.8825}   & 0.8858   & 0.8906\\
\cline{1-8}\\
& \multicolumn{7}{c}{\textbf{Loss}}\\ 
&\textbf{Mean} & \textbf{St.Dev}&  \textbf{q=0.025}&\textbf{q=0.25}&\textbf{q=0.5}&\textbf{q=0.75}&\textbf{q=0.975} \\ 
\cline{2-8}\\
Pairwise \textit{C}&    0.2483  &  0.0393   & 0.2200 &   0.2307   & \textbf{0.2409}  &  0.2518 &   0.2686\\
Pairwise \textit{SCR}&    0.1407 &   0.0050   & 0.1357  &  0.1388  &  \textbf{0.1405}   & 0.1426   & 0.1458 \\
\cline{1-8}
\end{tabular}
\end{scriptsize}
\end{center}
\end{table}

\newpage
\subsection{Data generated from a misspecified model}
\begin{table}[!htbp]
\begin{center}
\begin{scriptsize}
\caption{ Simulation results: ARI and loss for the posterior probabilities. Data generated
from a two-component  latent mixture; 5 ordinal variables with 5 categories; three of them are less informative. Separated groups. N=1000 and R=250 samples.}
\begin{tabular}{cccccccc}
\cline{1-8}\\
& \multicolumn{7}{c}{\textbf{Adjusted Rand Index}}\\ 
&\textbf{Mean} & \textbf{St.Dev}&  \textbf{q=0.025}&\textbf{q=0.25}&\textbf{q=0.5}&\textbf{q=0.75}&\textbf{q=0.975} \\ 
\cline{2-8}\\
Pairwise \textit{C}&  0.8970 &   0.1857 &   0.8470&    0.9392 &   \textbf{0.9672} &   0.9837   & 0.9918\\
Pairwise \textit{SCR}&  0.9950   & 0.0044   & 0.9918  &  0.9919  &  \textbf{0.9959}  &  0.9959   & 1.0000 \\
\cline{1-8}\\
& \multicolumn{7}{c}{\textbf{Loss}}\\ 
&\textbf{Mean} & \textbf{St.Dev}&  \textbf{q=0.025}&\textbf{q=0.25}&\textbf{q=0.5}&\textbf{q=0.75}&\textbf{q=0.975} \\ 
\cline{2-8}\\
Pairwise \textit{C}&   0.1152 &   0.1098   & 0.0360 &   0.0556   &\textbf{ 0.0839}  &  0.1155  &  0.1717 \\
Pairwise \textit{SCR}&     0.0269  &   0.0132   &  0.0116  &   0.0207    & \textbf{0.0282}    & 0.0327   &  0.0400   \\
\cline{1-8}
\end{tabular}
\end{scriptsize}
\end{center}
\end{table}

\begin{table}[!htbp]
\begin{center}
\begin{scriptsize}
\caption{ Simulation results: ARI and loss for the posterior probabilities. Data generated
from a two-component  latent mixture; 5 ordinal variables with 5 categories; three of them are less informative. Separated groups. N=5000 and R=250 samples.}
\begin{tabular}{cccccccc}
\cline{1-8}\\
& \multicolumn{7}{c}{\textbf{Adjusted Rand Index}}\\ 
&\textbf{Mean} & \textbf{St.Dev}&  \textbf{q=0.025}&\textbf{q=0.25}&\textbf{q=0.5}&\textbf{q=0.75}&\textbf{q=0.975} \\ 
\cline{2-8}\\
Pairwise \textit{C}& 0.9908 &   0.0087  &  0.9877 &   0.9918 &   \textbf{0.9934}  &  0.9943  &  0.9959 \\
Pairwise \textit{SCR}& 0.9954  &  0.0019 &   0.9934  &  0.9951  &  \textbf{0.9959}  &  0.9967 &   0.9975    \\
\cline{1-8}\\
& \multicolumn{7}{c}{\textbf{Loss}}\\ 
&\textbf{Mean} & \textbf{St.Dev}&  \textbf{q=0.025}&\textbf{q=0.25}&\textbf{q=0.5}&\textbf{q=0.75}&\textbf{q=0.975} \\ 
\cline{2-8}\\
Pairwise \textit{C}&   0.0381  &  0.0142 &   0.0274 &   0.0317   & \textbf{0.0355} &   0.0388  &  0.0473
\\
Pairwise \textit{SCR}&  0.0275 &   0.0053  &  0.0222   & 0.0253  &  \textbf{0.0275}  &  0.0300 &   0.0328 \\
\cline{1-8}
\end{tabular}
\end{scriptsize}
\end{center}
\end{table}

\begin{table}[!htbp]
\begin{center}
\begin{scriptsize}
\caption{ Simulation results: ARI and loss for the posterior probabilities. Data generated
from a two-component  latent mixture; 5 ordinal variables with 5 categories; three of them are less informative. Non-separated groups. N=1000 and R=250 samples.}
\begin{tabular}{cccccccc}
\cline{1-8}\\
& \multicolumn{7}{c}{\textbf{Adjusted Rand Index}}\\ 
&\textbf{Mean} & \textbf{St.Dev}&  \textbf{q=0.025}&\textbf{q=0.25}&\textbf{q=0.5}&\textbf{q=0.75}&\textbf{q=0.975} \\ 
\cline{2-8}\\
Pairwise \textit{C}&      0.4382   & 0.2819   & 0.0673  &  0.2324 &   \textbf{0.5563} &   0.6501 &   0.7172 \\
Pairwise \textit{SCR}&0.8817 &   0.0643  &  0.8571  &  0.8777  & \textbf{ 0.8915} &   0.9046   & 0.9165\\
\cline{1-8}\\
& \multicolumn{7}{c}{\textbf{Loss}}\\ 
&\textbf{Mean} & \textbf{St.Dev}&  \textbf{q=0.025}&\textbf{q=0.25}&\textbf{q=0.5}&\textbf{q=0.75}&\textbf{q=0.975} \\ 
\cline{2-8}\\
Pairwise \textit{C}&    0.3662 &   0.1273   & 0.2442  &  0.2752 &   \textbf{0.3200}  &  0.4323 &   0.5176 \\
Pairwise \textit{SCR}&  0.1497 &   0.0310   & 0.1280  &  0.1369  &  \textbf{0.1454 }  & 0.1546   & 0.1648 \\
\cline{1-8}
\end{tabular}
\end{scriptsize}
\end{center}
\end{table}

\begin{table}[!htbp]
\begin{center}
\begin{scriptsize}
\caption{ Simulation results: ARI and loss for the posterior probabilities. Data generated
from a two-component  latent mixture; 5 ordinal variables with 5 categories; three of them are less informative. Non-separated groups. Equidistant thresholds. N=5000 and R=250 samples.}
\begin{tabular}{cccccccc}
\cline{1-8}\\
& \multicolumn{7}{c}{\textbf{Adjusted Rand Index}}\\ 
&\textbf{Mean} & \textbf{St.Dev}&  \textbf{q=0.025}&\textbf{q=0.25}&\textbf{q=0.5}&\textbf{q=0.75}&\textbf{q=0.975} \\ 
\cline{2-8}\\
Pairwise \textit{C}&  0.5390  &  0.2498  &  0.1263 &   0.6519   & \textbf{0.6762}  &  0.6944   & 0.7083 \\
Pairwise \textit{SCR}& 0.9050  &  0.0114   & 0.8950 &   0.8997  &  \textbf{0.9055}  &  0.9102&    0.9161 \\
\cline{1-8}\\
& \multicolumn{7}{c}{\textbf{Loss}}\\ 
&\textbf{Mean} & \textbf{St.Dev}&  \textbf{q=0.025}&\textbf{q=0.25}&\textbf{q=0.5}&\textbf{q=0.75}&\textbf{q=0.975} \\ 
\cline{2-8}\\
Pairwise \textit{C}&   0.3176 &    0.0977   &  0.2466  &   0.2581  &   \textbf{0.2725}  &   0.2864   &  0.4607
\\
Pairwise \textit{SCR}&  0.1359   & 0.0076   & 0.1285 &   0.1324 &  \textbf{ 0.1358}  &  0.1390 &   0.1428  \\
\cline{1-8}
\end{tabular}
\end{scriptsize}
\end{center}
\end{table}

\newpage

\newpage

%

\end{document}